\documentclass[a4paper,12pt,titlepage]{article}
\usepackage{a4wide}
\usepackage{amsfonts}
\usepackage{amsmath}
\let\mathscr\mathcal
\usepackage{cite}

\def\N{\mathscr{N}}
\def\OO{\mathscr{O}}

\def\F{\mathcal{F}}

\def\X{\mathscr{X}}

\DeclareMathOperator{\tr}{tr}

\def\Im{\mathrm{Im}}
\def\Re{\mathrm{Re}}

\def\five-dimensional{5D}
\def\four-dimensional{4D}

\def\eqref#1{(\ref{#1})}

\def\AdS{\text{AdS}}
\def\CFT{\text{CFT}}
\def\Sp{Sp}
\begin{document}
\begin{titlepage}
\begin{flushright}
UT-05-08\\
hep-th/0507057
\end{flushright}
\vfil
\begin{center}
\LARGE Five-dimensional Supergravity Dual  of $a$-Maximization\\
\bigskip
\bigskip
\bigskip
\large {Yuji Tachikawa}\\
\bigskip
\textit {Department of Physics, Faculty of Science,}\\
\textit {University of Tokyo, Tokyo 113-0033,  JAPAN}\\
\bigskip
e-mail: \texttt{yujitach@hep-th.phys.s.u-tokyo.ac.jp}
\end{center}
\bigskip\bigskip\bigskip\bigskip\bigskip\bigskip
\centerline{\large\textbf{abstract}}
\bigskip
We study
 the five-dimensional
supergravity dual of the $a$-maximization
under $\AdS_5/\CFT_4$ duality.
We firstly show that
the $a$-maximization is mapped to
the attractor equation in five-dimensional gauged supergravity,
and that 
the trial $a$-function 
is the inverse cube of the
superpotential of the five-dimensional theory.

There is also a version of $a$-maximization in which
one extremizes over 
Lagrange multipliers enforcing the anomaly-free condition
of the R-symmetry.
We identify the supergravity dual of this procedure, and
show how the Lagrange multipliers
appearing in the supergravity description
naturally correspond to the gauge coupling of the superconformal field theory.

\bigskip\bigskip\bigskip

\end{titlepage}

\section{Introduction}

Conformal field theories (CFTs) are fascinating subjects in physics.
Often they are strongly coupled, which means the application of
na\"\i ve perturbation theory will be troublesome.
However, CFTs have more symmetries in addition to the usual
Poincar\'e symmetry.
Two-dimensional CFTs, in particular, have an infinite number of
symmetry generators, and the symmetry determines many of the
physical properties.  One can predict anomalous dimensions of various
operators, and they agree with those of critical phenomena
realized in nature.

CFTs in more than two dimensions remain more elusive.
Supersymmetric conformal field theories
(SCFTs) are somewhat better understood,
since 
the algebra relates the scaling dimension to the R-charge.
However, the precise understanding was limited, until quite recently,
to the models
with no global $U(1)$ symmetries other than the R-symmetry.
In a seminal  paper \cite{amax},
Intriligator and Wecht showed how to proceed
when there are several anomaly free R-currents.
It is found that the correct R-symmetry 
can be found by maximizing the so-called trial $a$-function.
This allowed various detailed study of previously unexplored SCFTs
and flows between them; see for example \cite{kp,iw}.
$a$-maximization has also found some phenomenological application
\cite{thatphenomenologypaper}.

Another way to approach the study of four dimensional CFT
is the use of Anti-de Sitter (AdS)/CFT duality \cite{Maldacena}. 
Indeed, using the prescription in
\cite{GKP,W}, any phenomena in CFT can be in principle mapped to
those in the gravitational theory on $\AdS_5$. 
Many candidates were constructed by using D3 brane
probing a Calabi-Yau cone with base $X_5$.
It predicts the duality between quiver gauge theories on the worldvolume of
the brane and the type IIB string on $\AdS_5\times X_5$.
One of the recent remarkable developments is 
that the gauge theory probing the tip of the
toric Calabi-Yau cone can be constructed
and analyzed by using the
$a$-maximization (see for example \cite{hanany}).
Another is that we saw a great advancement in our
understanding of the geometry of $X_5$ \cite{MSY}.
It has been discussed that the minimization of the volume of $X_5$ and the
maximization of $a$ always agree \cite{BZ}.

Much of the properties of the AdS side, however, can be studied without
recourse to the machinery of string theory.
We can 
consider the gravitational theory on $\AdS_5$ by itself.
The SCFT has at least eight supercharges, which means
the corresponding theory on $\AdS_5$ is
a five dimensional supergravity with
the same number of supercharges.
It is known that such theories have a quite rigid structure
which should be reflected on the properties of SCFTs.
The analysis of this relation
is performed
for the maximal supergravity 
in \cite{FerraraN4} and for
theories with sixteen and eight supersymmetries for example in
\cite{FerraraN1N2,holographicflow,T11sugraCFT}.
We think that it is worthwhile to
pursue this direction further,
by utilizing the lessons from the recent
advancement in understanding the SCFT in four dimensions.
It will enable us to use the intuition in one side to explore the other
in the AdS/CFT duality.

As the first step, we study the $\N=2$ supergravity dual
of the $a$-maximization of $\N=1$ SCFT.
We will find that the maximization procedure is mapped precisely
to the  attractor equation in supergravity.
We will also find the dual of the $a$-maximization
with the Lagrange multipliers introduced in \cite{kutasov,kutasovschwimmer}.
In that framework, anomalous and non-anomalous global symmetries
are treated on the same footing, and the anomaly-free condition for the
superconformal R-symmetry is imposed by the Lagrange multipliers.
It is proposed that these multipliers can be thought of as the
coupling constants of the gauge fields. We will see that
indeed there is a natural identification of the Lagrange multipliers
with the coupling constants from the supergravity point of view.

The organization of the paper is as follows:
we will first review in section 2
relevant aspects  of the gauged supergravity in five dimensions\cite{gunay,general}.
The discussions will be brief and used mainly to fix the notations.
Then in section 3, we show how the $a$-maximization is mapped
to the attractor equation in supergravity. We also study the supergravity
dual of the space of marginal deformations in SCFT.
In section 4,
we study the
$a$-function with Lagrange multipliers 
in the supergravity side.
We conclude in section 5 with some discussions and future prospects.

Note Added:  An interesting paper \cite{currentcorrelator} with
some overlapping material with ours
appeared on the eprint  archive a week after we posted ours there.

\section{Gauged supergravity in five dimensions}
Let us recall the structure of gauged supergravity in five dimensions.
The minimum number of supersymmetry generators 
is eight, and we concentrate on this case.
It is called $\N=2$ supergravity in the supergravity literature.
There are several kinds of supermultiplet,
and we restrict our attention to the gravity multiplet, vector
multiplets and hypermultiplets. We restrict attention to the
Abelian vector multiplets for brevity.
Non-abelian gauge fields can be incorporated without much effort.
We follow the conventions in \cite{general}.

\subsection{Structure of scalar manifolds}
Let us first discuss the scalars of the vector multiplets.
Let us denote the number of vector fields by $n_V$.
Then, there are $n_V-1$ real scalar fields in the theory.
When one compactifies one dimension, it will give
$\N=2$ supergravity in four dimensions. As such,
the structure of the scalar manifold is determined by a unique
function $\F$. A peculiarity in five dimension is that 
the third derivative of $\F$ governs the Chern-Simons coupling
and this fact fixes $\F$ to be cubic, \begin{equation}
\F=c_{IJK} h^Ih^Jh^K,\label{c}
\end{equation} where $h^I$, ($I=1,2,\ldots,n_V$), are
the special coordinates. The scalar manifold $M_V$ for the vector multiplet
is given by the real $n_V-1$ dimensional hypersurface 
defined by the constraint $\F=1$ in the space of $h^I$.
These manifolds are known as the
very special manifolds\cite{veryspecial}.
Let $\phi^x$, ($x=1,2,\ldots, n_V-1$),
parametrize the 
manifold. It is useful to introduce the following quantities:
\begin{equation}
h_I\equiv c_{IJK}h^Jh^K,\qquad
g_{xy}\equiv - 3c_{IJK}h^I_{,x} h^J_{,y} h^K, \qquad
a_{IJ}\equiv h_Ih_J+\frac32g^{xy}h_{I,x} h_{J,y}.\label{vectormetric}
\end{equation} We will raise and lower the indices $I, J, \ldots$ by using
$a_{IJ}$.

Let us turn to the hypermultiplets.
The manifold $M_H$ of the hyperscalars is
a quaternionic manifold 
of real dimension $4n_H$,
which means that its holonomy is contained in $\Sp(n_H)\times \Sp(1)$
\cite{BaggerWitten}.  
Let $q^X$, ($X=1,\ldots,4n_H$), parametrize the manifold. We will
introduce the vielbein $f^X_{iA}$ where $i=1,2$ and  $A=1,\ldots,2n_H$
are the indices for $\Sp(1)$ and $\Sp(n_H)$ respectively.
We normalize $f^X_{iA}$ so that $f_{iAX} f^{iA}_Y=g_{XY}$.
Supersymmetry 
fixes the $\Sp(1)$ part of the curvature \cite{BaggerWitten} so that
it is proportional to the triplet of almost complex structures:\begin{equation}
R_{XYij}=-(f_{XiC} f^C_{Yj}-f_{YiC}f_{Xj}^C).
\end{equation}  We trade the symmetric combination of two indices $\{ij\}$
for an index $r=1,2,3$ by using the Pauli matrices, that is,
\begin{equation}
T_{ij}=\sigma^r_{ij} T_r
\end{equation} for any tensors. The $\Sp(1)$
 curvature $R_{XY}^r$ satisfies the relation
\begin{equation}
R_{XY}^r R_{XZ}^s = \frac14\delta^{rs}g_{YZ}+\frac12 \epsilon^{rst} R^t_{YZ}.
\end{equation}

The kinetic terms for boson fields are then given by\begin{multline}
e^{-1}{\mathcal L}_{\text{kin,boson}}= -\frac12 R
 -\frac12 g_{xy} \partial_\mu \phi^x \partial^\mu \phi^y
  -\frac12 g_{XY} \partial_\mu q^X \partial^\mu q^Y\\
 -\frac14 a_{IJ} F^I_{\mu\nu} F^{J \mu\nu}
 +\frac1{6\sqrt{6}} e^{-1} c_{IJK} \epsilon^{\mu\nu\rho\sigma\tau}
 A^I_\mu F^J_{\nu\rho} F^K_{\sigma\tau} \label{kinetic}
\end{multline}where $R$ is the scalar curvature and $e$ is the determinant
of the f\"unfbein.

\subsection{Gauging and the Potential}
We need to introduce a scalar potential to get 
the $\AdS_5$ vacuum.
The structure of the potential 
is extremely restricted by the high degree of supersymmetries, and
it must be accompanied by the 
gauging of the scalars.
It also modifies the supersymmetry transformation. 

We will take isometries $K_I^X$ on $M_H$ to covariantize the
derivative \begin{equation}
\partial_\mu q^X\to  \partial_\mu q^X + A^I_\mu K_I^X \label{gauging}.
\end{equation} The isometries can be expressed by the relation
 \begin{equation}
 K_I^X R_{XY}^{ij}=D_Y P^{ij}_I,\label{hamiltonaction}
 \end{equation} using the Killing potential $P^{ij}_I$.
 This is required by the consistency of the gauging
 with the supersymmetry.
 
 The Killing potential $P^{ij}_I$ appears in the Lagrangian.
It gives the scalar potential as 
 \begin{equation}
V=\frac32g^{xy}\partial_x P^{ij}\partial_y P_{ij} +
\frac12 g^{XY}D_X P^{ij} D_Y P_{ij} - 2 P^{ij}P_{ij}\label{potential}
\end{equation}where $P_{ij}\equiv h^IP_{Iij}$.

It also appears in the covariant derivative of the gravitino $\psi_\mu^i$,
\begin{equation}
D_\nu \psi^i_{\mu}=\partial_\nu\psi^i_\mu+ A^I_\nu P^i_{jI} \psi^j_\mu+\cdots.
\label{R-gravity}
\end{equation} $P^i_{jI}$ enters in the covariant derivative of the gaugino as well.

It appears also in the supersymmetry transformation laws: \begin{align}
\delta_\epsilon \psi_{\mu}^i&= D_\mu \epsilon^i + \frac i{\sqrt{6}}\gamma_\mu
\epsilon^i P_{ij}+\cdots,\\
\delta_\epsilon \phi^x&=\frac{i}2 \bar\epsilon^i \lambda^x_i,\label{susytr-gauge}\\
\delta_\epsilon \lambda_x^i&= -\epsilon_j\sqrt{\frac23}\partial_x P^{ij}+\cdots\label{susytr-gaugino}\\
\delta_\epsilon q^X &= -i\bar\epsilon_i f^{XiA} \zeta_A,\label{susytr-hyper}\\
\delta_\epsilon \zeta_A &= \frac{\sqrt{6}}{4} \epsilon^i f_{XiA} K_I^X h^I +\cdots.
\label{susytr-hyperino}
\end{align}where $\lambda_x$ and $\zeta_A$ are
the gaugino and the hyperino, respectively.

Let us discuss a bit more about the isometry of the hyperscalars.
The relation \eqref{hamiltonaction} can be solved to give $P$ in terms of $K$
as follows:\begin{equation}
2n_H P^{ij}_I=D_X K_{I}^{Y} R_Y^{Xij}.
\end{equation}Consider a point on $M_H$
so that $K_I^X=0$, around which one can expand \begin{equation}
K_I^X=Q_{IY}^X q^Y + \OO(q^2).\label{chargeofhyper}
\end{equation} Comparing with \eqref{gauging}, we see
that $Q_{IY}^X$ determines the charge of the hypermultiplets.
Then $P^{ij}_I$ at the point is given by \begin{equation}
P^{ij}_I=\frac{1}{2n_H}Q_{IY}^X R_{X}^{Yij}.\label{Rpart}
\end{equation}It means that $P^{ij}_I$ is the $\Sp(1)$ part of the
charges $Q_{IY}^{X}$. Assuming $Q$s to be rational,
$P^{ij}_I$ at the point is also rational.

After this  review, we can now move on to
the study of the duality between $d=4$ $\N=1$ SCFT and the
$d=5$ $\N=2$ gauged supergravity.

\section{Dual of $a$-maximization}
\subsection{Brief review of $a$-maximization}
Let us consider an $\N=1$ SCFT in four dimensions.
Let us denote by $J_I^\mu$, ($I=1,2,\ldots,n_V$),
the currents of non-anomalous global symmetries of the theory
 and by $Q_I$ corresponding charges.
We demand the charges $Q_I$ to be integers, so that
$J_I$ can be coupled to external $U(1)^{n_V}$ connections
as follows:\begin{equation}
S\to S + J_I^\mu A^I_\mu+\cdots .
\end{equation} 
Some of them may 
rotate the supercoordinates $\theta_\alpha$.
Let the charges of $\theta_\alpha$ under the 
global symmetry be given by $\hat P_I$:\begin{equation}
\theta_\alpha\ \longrightarrow\ e^{i\phi^I Q_I}\theta_\alpha=
e^{i\phi^I\hat P_I}\theta_\alpha
\label{R-gauge}
\end{equation}
We will call a global symmetry $t^I J_I$ 
which commutes with $\theta_\alpha$ a flavor symmetry.
The condition is given by\begin{equation}
t^I \hat P_I=0\label{flavor}.
\end{equation}

Global symmetries, even if they are non-anomalous,
may have chiral anomalies among them.
This can be expressed by saying that the 
gauge transformation of the external $U(1)^N$ gauge field 
will have gauge anomaly described through the descent construction
by the anomaly polynomial \begin{equation}
\frac{1}{24\pi^2} \hat c_{IJK} F^IF^JF^K\label{anomalypolynomial}
\end{equation}where
$F^I=F^I_{\mu\nu}dx^\mu\wedge dx^\nu/2$
is the curvature two-form of the $I$-th external $U(1)$ gauge field.
The constants $\hat c_{IJK}$ are given by \begin{equation}
\hat c_{IJK}=\tr Q_I Q_J Q_K \label{c-hat}
\end{equation} where the trace is over the labels of the Weyl fermions of the theory.
There may be gravitational anomaly given by \begin{equation}
 \hat c_{I}F^I \tr R R
\end{equation}where $R$ is the curvature two-form of the external metric.

A particular combination of global symmetries
appears in the anticommutator of the supertranslation $Q_\alpha$ and
the special superconformal transformation $S^\alpha$ : \begin{equation}
\{Q_\alpha, S^\alpha \}\sim  \tilde s^I Q_I. \label{SCA}
\end{equation} We normalize $\tilde s^I$ so that
the charge of $\theta_\alpha$ under $\tilde s^I Q_I$ be one,
that is, $\tilde s^I \hat P_I=1$.
We denote  the superconformal R-symmetry by $R_{SC}=\tilde s^I Q_I$.

$R_{SC}$ can be used to uncover many physical properties
of the theory considered.  One is that
the scalar chiral primary will have dimension $\Delta$ given by
$\frac32 R_{SC}$. Another relation 
is with the central charges of the theory.
In four dimensions, there are two of them, $a$ and $c$, which
are defined as the coefficients in front of the Euler density and the 
square of the Weyl tensor in the trace anomaly of the theory.
They are expressible in terms of the superconformal R-symmetry as follows
\cite{anselmifreedman}:
\begin{equation}
a=\frac3{32}(3 \tr R_{SC}^3-\tr R_{SC}),\qquad
c=\frac1{32}(9\tr R_{SC}^3-5\tr R_{SC}).\label{fieldtheoryresults}
\end{equation}

Suppose that some high-energy description of the (possible) SCFT is given.
One can identify the non-anomalous symmetry and
can calculate $\hat c_{IJK}$ by using 't Hooft's anomaly matching.
The charges $\hat P_I$ of $\theta_\alpha$ under $J_I$
will also be easily given.
Then, the basic problem 
is the identification of the superconformal R-symmetry $R_{SC}=\tilde s^I Q_I$.

Here comes the brilliant idea of Intriligator and Wecht\cite{amax}. 
Let $Q_F=t^I Q_I$ be a flavor symmetry, i.e.
$t^I\hat P_I=0.$
They showed that
the triangle diagram with one $Q_F$ and two $R_{SC}$ insertions
can be mapped, by using the superconformal transformation,
to the triangle diagram with $Q_F$ and two energy-momentum tensor
insertions. The precise coefficient was calculated to give the relation 
\begin{equation}
	9\tr Q_F R_{SC} R_{SC} = \tr Q_F.\label{amax1}
\end{equation} Another requirement is the negative definiteness
\begin{equation}
	\tr Q_F Q_F R_{SC}<0.\label{amax2}
\end{equation}
Let us introduce
the trial $a$-function $a(s)$ for a trial R-charge $R(s)=s^\Lambda Q_\Lambda$
to be
\begin{equation}
a(s)=\frac3{32}(3\tr R(s)^3 -\tr R(s) ).
\end{equation}
The conditions \eqref{amax1}, \eqref{amax2} mean that $a(s)$
is locally maximized at the point
$s^I = \tilde s^I$, where the trial R-charge becomes the superconformal
R-charge $R_{SC}$.
It is understood that $s$ is constrained so that the charge of $\theta_\alpha$
under $R(s)$ is one.

\subsection{Supergravity dual of $a$-maximization}\label{dualamax}

We would like to see how the $a$-maximization is translated under the 
AdS/CFT duality to the supergravity description.
Let us first recall the prescription of AdS/CFT correspondence\cite{GKP,W}.
For a current $J_I^\mu$ in the CFT side, we introduce a gauge field
$A^I_\mu$ in the bulk which couples to the current
at the boundary with the interaction
$\int d^4x J_I^\mu A^I_\mu +\cdots$.
There are chiral anomalies among global symmetries generated by $J_I$,
which translates to the fact that
when the global symmetry is gauged, the partition function
depends on the gauge, see \eqref{anomalypolynomial}.  This can be reproduced 
through the bulk Chern-Simons
coupling \begin{equation}
\frac{1}{24\pi^2}\int \hat c_{IJK} A^I \wedge F^J \wedge F^K.
\end{equation}
It is gauge invariant on a manifold without boundary,
but it causes the partition function to vary appropriately
on a manifold with boundary.
This mechanism is the supergravity
realization of the descent construction of consistent
anomalies in the AdS/CFT correspondence\cite{W}.
Thus, we can identify the constants $\hat c_{IJK}$ in the CFT side \eqref{c-hat}
 and the constants $c_{IJK}$ in the AdS side \eqref{kinetic}:\begin{equation}
c_{IJK}=	\frac{\sqrt{6}}{16\pi^2}\hat c_{IJK}.
\end{equation}

Just in the same way, the chiral anomaly for the global symmetry--gravity--gravity
triangle diagram is reproduced by the coupling\begin{equation}
 \int \hat c_{I} A^I \wedge \tr R \wedge R
\end{equation}where $R$ is the curvature two-form of the metric.
It is, however, a higher derivative effect in the AdS side \cite{subleading}
so we 
neglect them in the rest of the paper. We hope to revisit the issue
in a later publication.
This means on the CFT side
that we restrict attention to theories
in which the chiral anomaly concerning gravity
is much smaller than the chiral anomaly among three $U(1)$ symmetries.

For the rest of the section let us assume that the hyperscalar is at the point
where $K_I^X=0$ and concentrate on the behavior of the vector multiplets.
Let us denote the charges of the hypermultiplet and the Killing potential
by $Q_{IY}^X$ and $P_I^{ij}=Q_{IY}^X R_{X}^{Yij}/(2n_H)$,
respectively, as in \eqref{chargeofhyper}.
The global structure of $M_H$  does not concern us\footnote{
For example, one can think of $M_H$ as one of the Wolf spaces such as
$\Sp(n_H,1)/\Sp(n_H)\times \Sp(1)$, and think of the Killing vectors
as induced by the subgroup of the denominator $\Sp(n_H)\times\Sp(1)$.
However, only the local properties of the metric
near the zero of  the Killing vectors
are relevant, as long as we restrict our attention to the charges
and the mass squared of the scalars as we will see below.
Another thing one should notice is that the hyperscalar contains the
dilaton, when the five-dimensional supergravity arises
as the compactification of the type IIB string theory. It means
that in general there are corrections to the metric of the hyperscalars.
Hence the final metric will not be as simple as the one for the Wolf spaces.}.
The global structure will be important if we study the flow
between two supersymmetric vacua \cite{ceresole}.

In order for the four-dimensional theory to be superconformal,
the five-dimensional bulk should be AdS, and
there should be eight covariantly constant spinors.
To achieve this, we need to set the gaugini variation \eqref{susytr-gaugino}
to be zero, \begin{equation}
P_I^{r} \tilde h^I_{,x} =0\label{gauginovar0}.
\end{equation}
We denoted the value of the quantity 
at the AdS vacuum by adding a tilde.
This condition says that the three vectors $P_I^r$, $r=1,2,3$,
is perpendicular to the $n_V-1$ row vectors $\tilde h^I_{,x}$
as vectors with  $n_V$ columns, which in turn means
that  $P_I^r$ are parallel.
Thus we can use the $\Sp(1)$ global R-symmetry
to set $ 
P^r_I = \delta^{3r} P_I
$ 
for some constants $P_I$.
Then the equation \eqref{gauginovar0} reduces to\begin{equation}
P_I \tilde h^I_{,x}=0.\label{gauginovar}
\end{equation}
This is  an extremization condition 
for the superpotential $P\equiv P_I h^I$.

Now one can determine the commutation relation among the
global symmetries which are respected by the vacuum.
They are the isometries of $\AdS_5$, eight supercharges,
and $n_V$ global $U(1)$ symmetries
from the gauge fields $A_\mu^I$.
From the covariant derivative of the gravitino \eqref{R-gravity},
one finds that supercharges have charge $\pm P_I$ under the $I$-th global
$U(1)$.   Thus we can identify the quantity $P_I$ introduced above
and the quantity $\hat P_I$ in the SCFT side which was introduced in
\eqref{R-gauge}.  We will not distinguish $P_I$ and $\hat P_I$ in the following.

We have found the mapping under AdS/CFT duality
of the basic constants $\hat c_{IJK}$ and $\hat P_I$ in the SCFT
and $c_{IJK}$ and $P_I$ in the supergravity.
Now we can study how the $a$-maximization is translated 
on the gravity side.
Let us resume the study of the
implication of the condition \eqref{gauginovar}
and recall the constraint $c_{IJK}h^Ih^Jh^K=1$,
which implies that $ h_I h^I{}_{,x} = c_{IJK} h^I h^J h^K_{, x} = 0$.
Thus $h_I$ also is perpendicular to $n_V-1$ vectors
$h^I_{,x}$, from which we deduce that \begin{equation}
\tilde h_I=c_{IJK}\tilde h^J\tilde h^K \propto P_I. \label{attractor}
\end{equation}This is the attractor equation in the five-dimensional
gauged supergravity\cite{soo-jong}.

Let us now identify the superconformal R-symmetry $R_{SC}=\tilde s^I Q_I$.
From the supersymmetry transformation law for the hypermultiplets
\eqref{susytr-hyper}, \eqref{susytr-hyperino}, 
we can calculate the anticommutator of the supercharges
acting on the hyperscalars. The result is
\begin{equation}
\{\delta_\epsilon,\delta_{\epsilon'}\} q^X = -i\frac{\sqrt{6}}4 (\bar\epsilon\epsilon')
\tilde h^I K_I^X +\cdots, 
\end{equation}from which we deduce that the anticommutator of the supercharges
contains a $U(1)$ rotation $\propto \tilde h^I Q_I$.
This $U(1)$ symmetry is identified under the AdS/CFT duality
with the $U(1)_R$ symmetry in the superconformal algebra \eqref{SCA}.
Thus we find \begin{equation}
\tilde s^I= t \tilde h^I. \label{prop}
\end{equation}  where $t$ is some proportionality constant.
Let us next fix $t$.
The gauge transformation law for the gravitino \eqref{R-gravity}
signifies that the superconformal R-charge of  the gravitino is
$\tilde s^I P_I$.
Considering that the superconformal R-symmetry is defined to rotate the
gravitino by charge one, we need $\tilde s^I P_I=1$.
Thus we get \begin{equation}
\tilde s^I=\tilde h^I/\tilde P,\label{yey}
\end{equation}where $\tilde P=\tilde h^I P_I$.
Recall that a flavor symmetry $t^I Q_I$ satisfies $P_I t^I=0$, see \eqref{flavor}.
Plugging this into the attractor equation \eqref{attractor},
we obtain\begin{equation}
c_{IJK}\tilde h^J\tilde h^K  t^I \propto P_I t^I =0.
\end{equation} This is precisely the condition \eqref{amax1}
for theories with no chiral anomaly concerning gravity.

The other equation \eqref{amax2} is, by using \eqref{vectormetric},
translated to the positivity of the metric of the scalar manifold \cite{ritz}.
To see this, let us recall that the $n_V-1$ vectors $\tilde h^I_{,x}$ spans 
the vector space $F$ defined by the condition \begin{equation}
F=\{t^I \ | \ P_I t^I=0\}.
\end{equation}
Thus the positivity of the matrices $-\hat c_{IJK}\tilde s^I$
acting on $F$,
equation \eqref{amax2}, is translated to the positivity of 
the matrix $-\hat c_{IJK}\tilde s^I \tilde h^J_{,x}\tilde h^K_{,y}$,
which is precisely the metric \eqref{vectormetric}
of the vector multiplet scalars.

The maximization of the trial $a$-function $a(s^I)$ and the extremization
of the superpotential $P=P(h^I)$ can be associated more explicitly. 
Let us generalize the relation \eqref{yey} and relate the parameter for the
trial R-symmetry $s^I Q_I$ and the value of the special coordinates $h^I$
by the formula $s^I=h^I/(P_J h^J)$.
Then,  we have\begin{equation}
a(s)\propto c_{IJK} s^Is^Js^K=\frac{c_{IJK}h^Ih^Jh^K}{(P_Ih^I)^3}=(P_I h^I)^{-3}.
\end{equation}
Thus, the trial $a$-function of the SCFT
is precisely the inverse cube of the superpotential.
Now it is trivial to see that the minimization of $a$ is the maximization of $P$ !

Let us carry out another 
consistency check.    It is known \cite{holographicWeyl,Gubser,NOO}
that, in a five-dimensional gravitational theory
with the action \begin{equation}
S=\frac12\int d^5x \sqrt{g}(R+12\Lambda +\cdots),
\end{equation} the central charge $a$ is given by \begin{equation}
a=\pi^2 \Lambda^{-3/2}.\label{a-lambda-relation}
\end{equation}
At the AdS vacuum,  the negative of the vacuum energy 
is given from the potential \eqref{potential} so that\begin{equation}
6\Lambda=-V=4(P_I\tilde h^I)^2.\label{lambda-P}
\end{equation}
Plugging the relation \eqref{yey} and \eqref{lambda-P}
into \eqref{a-lambda-relation}, we get
\begin{equation}
a=\pi^2 \left(\frac32\right)^{3/2}c_{IJK}\tilde s^I \tilde s^J \tilde s^K
=\frac9{32}\hat c_{IJK}\tilde s^I\tilde s^J\tilde s^K. 
\end{equation}
This agrees with the result from the field theory \eqref{fieldtheoryresults}.

\subsection{Mass squared of the vector multiplet scalar}
The result  presented in this and the next subsections is not new,
see for example \cite{holographicflow}. 
It is a preparation for section \ref{marginalsec} and section \ref{lagrange}.

We would like to study next the behavior of
scalars in the vector multiplet  around the vacuum $\tilde h^I$.
We can calculate the second derivative of the potential there
by using the special geometry relation\begin{equation}
h^I_{,x;y}=\frac{2}{3}h^Ig_{xy}+T_{xyz}h^{I}_{,w}g^{zw}
\end{equation}where $T_{xyz}$ is a certain completely symmetric tensor
on $M_V$. Then,
\begin{equation}
D_x \partial_y V \bigm|_{h^I=\tilde h^I}=-g_{xy}\frac83\tilde P^2
\end{equation} 
Thus,  the mass squared for all the scalar fields is negative
with $m^2=-4\Lambda$.
Recall  the classic relation \begin{equation}
m^2/\Lambda=\Delta(\Delta-4)
\end{equation}between the mass $m$ in the AdS and the dimension
$\Delta$ in the CFT.
Then we have  $\Delta=2$ for all the $N_V-1$ scalar fields,
which barely saturates the Breitenlohner--Freedman bound\cite{BrFr}
and thus the system is stable \cite{townsend}.
It is also easy to see that they have no R-charges.
This is as it should be, because the scalar component of
a vector multiplet corresponds to the lowest component of the
current superfield, whose dimension is protected and whose R-charge is zero.

\subsection{Dual of Scalar Chiral Primaries in SCFT}
\label{chiralprimary}
An important kind of multiplets in the $d=4$ $\N=1$ SCFT is
the chiral multiplet, whose lowest component is a complex scalar
we denote by $\OO$.
From the superconformal algebra the dimension and the R-charge
is related through $\Delta(\OO)=3R(\OO)/2$.
We would like to identify its supergravity dual.
A natural candidate will be a hypermultiplet, which comes in
quartets of real scalars. 
Chiral primaries, however, comes in pairs of real scalars. 
Na\"\i vely there is twice the number of freedom in supergravity.
We will see shortly below that the extra two degree of freedom
corresponds to the F-component of the superfield $\OO$.

Consider $n_H$ hypermultiplets $q^X$ with charges $Q_{IY}^{X}$
under $I$-th $U(1)$ symmetry,
where $X,Y=1,\ldots,4n_H$.
The charges under the superconformal R symmetry are given by\begin{equation}
Q_Y^X\equiv \tilde s^I Q_{IY}^X=\tilde h^I Q_{IY}^X / \tilde P.
\end{equation}
First, we would like 
to study the eigenvalues of $Q_Y^X$.
To facilitate the task, let us combine the $4n_H$ real scalars
into $2n_H$ complex scalars by introducing some complex structure
$J_X^Y$ so that  $Q_Y^X$ is diagonal. The relation \eqref{Rpart} between
$P^{r}$ and $Q_Y^X$ means that $J_X^Y$ is proportional to
$R_{X}^{Yr}P_{r}$.
Let us form $J^\pm$ from the other two complex structures
so that \begin{equation}
[J,J^\pm]=\pm 2 J^\pm.
\end{equation}
Then we can calculate the commutation relations of $Q_Y^X$ and 
three $J$'s with the results\begin{equation}
[J,Q]=0,\qquad
[Q,J^\pm]=\pm 2 J^\pm.
\end{equation}
This means that 
the $2n_H$ eigenvectors can be arranged in pairs
$q_{iA}$ with charges $r_{iA}$ , ($i=1,2$ and $A=1,\ldots,n_H$), 
so that $r_{1A}=r_{2A}+2$. We further abbreviate so that $r_A\equiv r_{1A}$.
One can also check that the supersymmetry 
relates $q_{1A}$ and $q_{2A}$.

The mass squared $m^2_{iA}$ for the scalar $q_{iA}$
can then be read off from the second derivative of the
scalar potential: 
\begin{align}
D_X\partial_Y V|_{h^I=\tilde h^I}&=g^{ZW}D_XD_Z P^{ij} D_YD_W P_{ij}
-4P_{ij}D_XD_YP^{ij} \nonumber \\
&=\frac32 Q_X^Z Q_{YZ} \tilde P^2 -4 \tilde P^{ij}R^{ij}_{XZ}Q_Y^Z.
\end{align} Substituting the diagonalized form of the charge matrix,
we obtain the masses of the hypermultiplets as  follows:
\begin{align}
m^2_{1A}
&=\frac32r_A(\frac32r_A-4)\Lambda,\\
m^2_{2A}
&=(\frac32r_A+1)(\frac32r_A-3)\Lambda.
\end{align}

Thus, the scalar which is dual to $q_{1A}$ under AdS/CFT has
dimension $3r_A/2$ and R-charge $r_A$, and the one for $q_{2A}$
has dimension $3r_A/2+1$ and R-charge $r_A-2$.
This combination of dimensions and charges are precisely the ones for
the lowest component and the F component of a chiral multiplet.

\subsection{Dual of Marginal Deformations}\label{marginalsec}
Let us next discuss the supergravity
dual of exactly marginal deformations in SCFT,\begin{equation}
S\to S+\int d^4x d^2\theta \tau_i \OO_i
\end{equation} where the  superconformal R-charge of the
operators $\OO_i$ should be two.
As remarked in \cite{conformalmanifold}, 
$\tau_i$ form a manifold
$M_c$ which parametrize the finite deformation.
$M_c$ naturally has a complex structure on it,
which comes from  the fact that the superpotential terms in $d=4$ $\N=1$
field theories have natural holomorphic structure.

We should be able to identify $M_c$ in the framework of supergravity.
As found in the last subsection, 
infinitesimal deformations
with chiral primaries correspond to the hypermultiplet scalars.
That the R-charge of a chiral primary is two means that the mass squared of the
corresponding hypermultiplet  scalar is $0$ and $-3$, and we saw
the deformation $\int d^2\theta \tau_i \OO_i=\tau_i [\OO_i]_F$ 
corresponds to the two real scalars of mass squared zero.
This in turn signifies that, when there are $n$ chiral primaries of
R-charge two, the supergravity vacuum comes in families with
$2n$ real parameters. Let us call it $M_{\text{sugra}}^c$.
This should be the supergravity realization of $M_c$.
From the supersymmetry transformation law,  we see that
\begin{equation}
M_{\text{sugra}}^c=\{  p\in M_H  \ |  \  \tilde h^I K_I^X(p)=0\}.
\end{equation}

One thing is not obvious, however. The hypermultiplet scalars 
form a quaternionic manifold, which is definitely not a complex manifold.
It is because the almost complex structures of a quaternionic manifold
is not  closed, but \textit{covariantly} closed.  Fortunately, we can easily
check that the submanifold $M_{\text{sugra}}^c$ is a K\"ahler manifold as follows.

First define $K^X\equiv \tilde h^I K_I^X$ and
$P^r\equiv\tilde h^I P_I^r$ for brevity.
From the property of the Killing potential $ D_X P^r = R^r_{XY} K^Y$,
$P^r$ is covariantly constant on $M_{\text{sugra}}^c$.
In particular, $P\equiv|P^r|$ is a constant parameter.
Thus, $J^X_Y$ defined by \begin{equation}
J^X_Y \equiv R_{YZ}^{r} g^{XZ} P_r/P \label{inducedcomplexstructure}
\end{equation} is an almost complex structure.
This $J^X_Y$ is covariantly constant with respect to the metric $g_{XY}$
restricted from $M_{H}$ onto $M_{\text{sugra}}^c$, because every factor in
\eqref{inducedcomplexstructure} is covariantly constant.
It tells us that the metric on $M_{\text{sugra}}^c$ has $U(n)$ holonomy,
which means that $M_{\text{sugra}}^c$ is K\"ahler.

\section{Dual of $a$-maximization with Lagrange multipliers}\label{lagrange}
\subsection{$a$-maximization in the presence of anomalous currents}
We saw in the previous section how the superconformal R-symmetry
can be found as the combination of non-anomalous global currents,
both from the SCFT and from the supergravity point of view.
In \cite{kutasov,kutasovschwimmer},
Kutasov and collaborators  incorporated anomalous global currents
to the picture.
It starts with the same trial $a$-function \begin{equation}
a(s)=\frac{3}{32}(3\tr (s^I Q_I)^3-\tr s^I Q_I ),
\end{equation}where $Q_I$ now include all the global symmetries,
anomalous or non-anomalous.  
Let us denote the $a$-th gauge fields by $F^a_{\mu\nu}$
and the Adler-Bell-Jackiw
anomaly coefficient of the $I$-th global symmetry with $a$-th gauge field
by $m_I^a$. 

We need to impose the 
anomaly-free condition for each gauge group in the field theory
considered.  Thus we have to introduce the Lagrange multipliers $\lambda_a$
and consider \begin{equation}
a(s,\lambda)=a(s)+\lambda_a (m_I^a s^I).\label{multipliergauge}
\end{equation} We need to extremize it with respect to both
$s^I$ and $\lambda_a$.
Define the function $a(\lambda)$ by first maximizing $a(s,\lambda)$ 
with respect to $s^I$, fixing $\lambda_a$.
When $\lambda=0$, the anomaly free condition is not imposed,
and the $a$-function takes the value for the free field theory.
This corresponds to zero gauge coupling. When $\lambda$ attains the value
$\tilde\lambda$
where $a(\lambda=\tilde\lambda)$ is extremized,
$a$ becomes the true central charge
of the SCFT.  In \cite{kutasov} it was shown that
$a$ generically decreases along the flow of $\lambda$ from zero to $\tilde \lambda$,
suggesting that $\lambda$ and the gauge coupling constants can be 
somehow identified.
Indeed,  there is a certain
renormalization scheme where such identification is precise
\cite{kutasovschwimmer}.
We would like to understand the supergravity dual of this procedure.
We will see that in the framework of five-dimensional supergravity,
the Lagrange multipliers and the gauge coupling constants
can be naturally identified.

In \cite{kutasov,kutasovschwimmer} Lagrange multipliers
were also introduced
for the condition that the superpotential
should have R-charge two. 
Since the analysis for that case is basically identical,
we concentrate on the interpretation of the Lagrange multipliers
for the anomaly-free condition.

\subsection{Supergravity Dual}
First we need to study the supergravity dual for the anomalous
global currents in SCFT.
Let us first discuss  without reference to supersymmetry.
The conservation law is modified by the anomaly to be \begin{equation}
\partial_\mu J^\mu = \X \label{anomalous}
\end{equation}for a suitable operator $\X$.
An example of such current is a chiral $U(1)$ rotation which is
broken by the instantons with $\X\propto \tr F\wedge F$.
Here $F$ is the curvature of the gauge field which is not external,
but is the constituent of  the CFT considered.

In order to consider the gravity dual, we introduce
external fields $A_\mu$ and $\phi$ defined on the AdS
and the coupling  $\int dx^4 A_\mu J^\mu  + \phi \X$
on the boundary.  We can see now  \cite{ouyang}
that if the bulk gauge transformation
$A_\mu\to A_\mu + \partial_\mu \epsilon$ is accompanied
by the transformation $\phi\to \phi-\epsilon$,
the prescription of AdS/CFT correspondence leads to the 
anomalous conservation law \eqref{anomalous}.

Let us now consider the effect of supersymmetry.
Consider an anomalous flavor symmetry.
The current $J^\mu$ is then incorporated into a current superfield $J$
and the operator $\X$ is the imaginary part of the F-component of the 
superfield $\OO\propto \tr W_\alpha W^\alpha$. The supersymmetry
completion of \eqref{anomalous}
becomes\begin{equation}
\bar D^2  J =\OO.
\end{equation} This is the celebrated Konishi anomaly \cite{KonishiShizuya}.

We saw that a current superfield in SCFT corresponds to a vector multiplet
and that a chiral multiplet to a hypermultiplet.
Thus, the Konishi anomaly 
is dual in the supergravity description
to the Higgsing of a vector multiplet eating a hypermultiplet.
After Higgsing, the multiplet is no longer short. Thus the dimension of
the operators is not protected anymore. However supersymmetry relates
the anomalous dimension of $J$ and the anomalous dimension of $\OO$ \cite{anselmi}.

Let us examine how the incorporation of these massive vector multiplets
is reflected to the $a$-maximization.
Consider the chiral operators $\OO_a=\tr W^\alpha_a W_{a \alpha}$
which yield kinetic terms for the $a$-th non-Abelian gauge fields,
where $a=1,\ldots,n_H'$ label the factor of the gauge groups.
We do not sum over $a$ inside the trace.
Define the isometries $K_a^X$ so that, if the Konishi anomaly for
the $I$-th current superfield is given by \begin{equation}
\bar D^2 J_I \propto m_I^a\OO_a,
\end{equation} the vector field $A^I_\mu$ in the five-dimensional
supergravity gauges the direction $m_I^aK_a^{X}$. $K_a^X$ is non-zero
at the vacuum.

Other $n_H''$ hypermultiplets which are not Higgsed are also charged
under $A^I_\mu$. We denote the Killing vectors for these
by $K_{(0)I}^{X}$. We assume that this can be expanded as
\begin{equation}
K_{(0)I}^{X}=Q^X_{IY} q^Y +\OO(q^2)\label{mmm}
\end{equation} as before.
The total gauging $K_I^X$
appearing in the supergravity Lagrangian is given by \begin{equation}
K_I^X=m_I^aK_a^{X}+K_{(0)I}^{X},
\end{equation}and we denote the corresponding Killing potential
as \begin{equation}
P_I^r=m_I^a P_a^r+P_{(0)I}^r.\label{Kp}
\end{equation} 

Let us study the condition for the AdS vacuum,
which can be found by inspecting the hyperino and gaugino transformation laws.
A convenient parametrization of the hyperscalars near the vacuum
is given as follows: near the zero of $K_{(0)I}^X$,
let it be linearly dependent on the scalars $q^{\hat X}$
where $\hat X=1,\ldots,4n_H''$.  We need $4n_H'$ coordinates in addition.
$n_H'$ of them are the gauge orbits along $K_a^X$.
We can take $P^r_a$ 
as the remaining $3n_H'$ of the coordinates.
They are guaranteed to form  good local coordinates because\begin{equation}
D_X P^{ij}_a = R_{XY}^{ij} K_a^X\ne 0.
\end{equation}

The hyperino transformation law gives the condition\begin{equation}
 \tilde h^I K_I^X=0.\label{first}
\end{equation} Its first consequence is that \begin{equation}
\tilde h^I m_I^a=0.
\end{equation}
Recall the superconformal R-charge is proportional to $\tilde h^I Q_I$,
see \eqref{yey}.
This translates in the SCFT language to the fact that
anomalous global currents do not participate in the superconformal R-charge.
Assuming other linear combination of $\tilde h^I$ is non-zero,
we can see that $q^{\hat X}=0$. 
However, hyperino variation alone does not fix $P^r_a$.

Next, let us turn to the gaugino variation \begin{equation}
\tilde h^I_{,x} P_I^{ij}=0.\label{second}
\end{equation} Just as before, it says that the  vectors  with $n_V$ elements
$P_I^{r=1,2,3}$ are all parallel to $h_I$. 
Global $\Sp(1)$ rotation can be used so that $P_I^r$ is nonzero only for
$r=3$.
Let us note that, from the relation \eqref{mmm}, $P_{(0)I}^{r=1,2}$ is 
quadratic in the fields $q^{\hat X}$.
Combining with the equation \eqref{Kp},
 we get $P^{r=1,2}_a=0$.
Thus, the remaining variables are $(n_V-1)$ vector multiplet scalars
and $n_H'$ coordinates of the hyperscalar, $P^{r=3}_a$.
Now define the superpotential to be the gravitino variation
\begin{equation}
P\equiv h^I P_{(0)I}+ P^{r=3}_am_I^ah^I \label{multipliergrav}
\end{equation} where the parameter is the vector multiplet scalars
$h^I$ with the constraint $c_{IJK}h^Ih^Jh^K=1$ and 
the hypermultiplet scalars $P^{r=3}_a$.
Extremization condition for those scalars yields precisely
the conditions \eqref{first} and \eqref{second}.
Thus, the AdS vacua can be found by extremizing the 
superpotential $P$ with respect to $h^I$ and $P^{r=3}_a$.
We can see that the scalars $P^{r=3}_a$ work as the Lagrange multipliers
enforcing the condition $h^I m_I^a=0$. 
Surprisingly, the Lagrange multipliers are physical fields
 in the supergravity side!

The way of introducing multipliers in the
gauge theory \eqref{multipliergauge} and in the supergravity
\eqref{multipliergrav} is not exactly the same.  We know that however,
when we want
to extremize a quantity, say $a(x)$, with respect to $x$ 
in the presence of the constraint $c(x)=0$,
it is immaterial whether we choose
$a_1(x,\lambda)\equiv a(x)+\lambda c(x)$ or
$a_2(x,\lambda)\equiv f^{-1}\left(f(a(x))+\lambda c(x)\right)$.
The difference between \eqref{multipliergauge} and \eqref{multipliergrav}
is of this form, thus of no physical relevance.

We have seen in section 3 that $(n_V-1)$ vector multiplet scalars
corresponds to the trial R-charge through the relation \eqref{yey},
and that the dual SCFT operator  is the scalar in the current superfield.
Then it is natural to ask the same question on the scalars $P^{r=3}_a$.
It acted as the Lagrange multipliers in the superpotential
extremalization. 
As a final exercise in this paper,
let us identify the SCFT dual of the scalar $P^{r=3}_a$.
The fact that $\tilde h^I P^r_I$ is nonzero only for $r=3$ means
that the superconformal R-symmetry is the $U(1)$ subgroup specified by
$\sigma^3$ of the global $\Sp(1)$ R-symmetry of the ungauged theory.
This can be read off just as in the discussion in the section \ref{dualamax}.
This tells us that $P^{r=3}_a$ has zero superconformal R-charge,
while $P^{r=1,2}_a$ has charge two.

We have already seen that the gauge orbit $K_a^X$  corresponds to the
topological density $\tr F^a\wedge F^a$.  Three real scalars $P^r_a$ are
its supersymmetric partners. 
From the discussions in section \ref{chiralprimary},
we can infer that $P^r_a$ correspond under AdS/CFT duality to the
three operators \begin{equation}
\tr F^a_{\mu\nu} F^a_{\mu\nu}, \qquad
\Re \tr\lambda^a_\alpha \lambda^{a\alpha}, 
\quad \text{and}\quad
\Im \tr\lambda^a_\alpha \lambda^{a\alpha}.
\end{equation}
Comparing the R-charges, we find that
$P^{r=1,2}_a$ are the dual  for the gaugino bilinears
and that  $P^{r=3}_a$
corresponds to the kinetic term $\tr F^a_{\mu\nu} F^a_{\mu\nu}$.
Let us recall that the prescription of AdS/CFT duality \cite{GKP,W}  means that
there is a boundary interaction \begin{equation}
\int dx^4 P^{r=3}_a \tr F^a_{\mu\nu} F^a_{\mu\nu}.
\end{equation}
Thus, we found that the Lagrange multiplier $P^{r=3}_a$
corresponds precisely to the gauge coupling constant under AdS/CFT duality.

\section{Conclusion and Discussions}
In this paper, we showed how the $a$-maximization in the 
SCFT is mapped to the attractor equation in five-dimensional
supergravity.  We saw that the lowest derivative terms of the 
gravitational theory is determined by the very quantities
$\hat c_{IJK}$ and $\hat P_{I}$
which enters the $a$-maximization.
We also saw how it is 
related to the attractor equation thanks to the structure
of the very special structure of the scalar manifold.
Furthermore, we studied the supergravity dual of 
another version of the $a$-maximization, which is
formulated as the 
extremization over the Lagrange multipliers enforcing the 
anomaly free condition. We showed that
the Lagrange multipliers also appear in the supergravity description,
and that they naturally corresponded to the
gauge coupling constant. This agrees with the expectations from the 
analysis in four-dimensional field theory.

We would like to discuss about the possible directions of research.
An immediate concern is the determination of $c_{IJK}$
when the five-dimensional theory is obtained as the compactification of
the type IIB superstring on a Sasaki-Einstein $X_5$.
For the CY compactification of M-theory, we can easily show
that $c_{IJK}$ is none other than the
triple intersection product of the two-cycles in the CY.
We could expect some kind of topological formula for
$c_{IJK}$ given the topology of $X_5$,
especially if $X_5$ is toric.

There also are several questions purely within the realm of
correspondence between
five-dimensional supergravities and four-dimensional SCFTs.
First is the incorporation of $A^I\wedge\tr R\wedge R$ term in the
supergravity. This will require the construction of five-dimensional
gauged supergravity with higher derivative terms.
Second is the study of the dual of the Higgsing in the SCFT.
SCFT can often be deformed by giving non-zero vacuum expectation
values to chiral primaries, and the moduli of such deformations
will be a K\"ahler cone. It will be extremely interesting if we can realize
this K\"ahler cone from the scalar manifold of supergravity.
It may be possible to phrase the result in \cite{KWHiggs}
in some purely five-dimensional parlance.
We would like to visit  these issues in future publications.

\paragraph{Acknowledgement} 
The author is greatly indebted to Y. Ookouchi for first bringing his attention to
this subject.   This short note emerged from the conversations with him.
The author also thanks to T. Eguchi, Y. Nakayama and T. Kawano
for useful discussions. The author is supported
by  Japanese Society for the Promotion of Science.

\end{document}